\documentclass[aps,prb,reprint,twocolumn,notitlepage,superscriptaddress, preprintnumbers]{revtex4-2}
\usepackage{graphicx}
\usepackage{amsmath,amssymb}
\usepackage{color}
\usepackage{bm}
\usepackage{bbm}
\usepackage{mathtools, physics}
\usepackage{dsfont}
\usepackage{hyperref}
\usepackage{amsthm, times}
\theoremstyle{plain}

\newtheorem*{thm*}{Theorem}

\begin{document}

\title{Frustration-free free fermions}
\author{Seishiro Ono}
\affiliation{Institute for Solid State Physics, University of Tokyo, Kashiwa 277-8581, Japan}
\affiliation{Department of Physics, Hong Kong University of Science and Technology, Clear Water Bay, Hong Kong SAR, China}
\affiliation{RIKEN Center for Interdisciplinary Theoretical and Mathematical Sciences (iTHEMS), RIKEN, Wako 351-0198, Japan}
\author{Rintaro Masaoka}
\affiliation{Department of Applied Physics, The University of Tokyo, Tokyo 113-8656, Japan}
\author{Haruki Watanabe}\email{hwatanabe@g.ecc.u-tokyo.ac.jp}
\affiliation{Department of Applied Physics, The University of Tokyo, Tokyo 113-8656, Japan}
\author{Hoi Chun Po}
\email{hcpo@ust.hk}
\affiliation{Department of Physics, Hong Kong University of Science and Technology, Clear Water Bay, Hong Kong, China}
\affiliation{Center for Theoretical Condensed Matter Physics, \\
	Hong Kong University of Science and Technology, Clear Water Bay, Hong Kong, China}

\date{\today}

\preprint{RIKEN-iTHEMS-Report-25}

\begin{abstract}
We develop a general theory of frustration-free free-fermion systems and derive the necessary and sufficient conditions for such Hamiltonians. Assuming locality and translation invariance, we find that any band touching between the valence and the conduction bands is always quadratic or softer, which rules out the possibility of describing Dirac and Weyl semimetals using frustration-free local Hamiltonians. We further construct a frustration-free free-fermion model on the honeycomb lattice and show that its density fluctuations acquire an anomalous gap originating from the diverging quantum metric associated with the quadratic band-touching points. Nevertheless, an $O(1/L^2)$ finite-size scaling of the charge-neutral excitation gap can be verified even in the presence of interactions, consistent with the more general results we derive in an accompanying work \href{https://arxiv.org/abs/2503.12879}{[arXiv:2503.12879]}.
\end{abstract}

\maketitle

{\it Introduction}---Frustration is a common enabler for exotic quantum phases of matter. In lattices like the kagome and pyrochlore, geometric frustration in the nearest-neighbor models results in destructive interference which quenches the kinetic energy of the electrons and leads to completely flat bands. In spin liquids, energetic frustration disfavors conventional magnetic orders and leads to highly entangled quantum disordered ground states hosting fractionalized excitations like anyons. 

More technically, we say a Hamiltonian is frustration-free if all the local terms are simultaneously minimized by the ground states~\cite{TasakiBook}. Frustration-freeness serves as a powerful starting point for deriving general results about both the ground state and the excitation spectrum of the system---a rare possibility for quantum many-body problems.
Restrictive as it may seem, a surprisingly diverse array of quantum phases can be realized using frustration-free Hamiltonians. Many important spin models are frustration-free, such as the Affleck--Kennedy--Lieb--Tasaki model~\cite{AKLT, TasakiBook}, the Majumdar--Ghosh (MG) model~\cite{10.1063/1.1664978, Majumdar:1970aa}, and more generally the parent Hamiltonians of matrix product states~\cite{parent, parent_PEPS}. 

Paradoxical as it may seem, frustration-free models also serve as important theoretical tools for investigating exotic quantum phases associated with geometric frustration.
Examples include the Rokhsar-Kivelson model~\cite{PhysRevLett.61.2376} for quantum critical points and the Kitaev toric code models~\cite{KITAEV20032} for quantum spin liquids, as well as the free-fermion nearest-neighbor hopping models on the kagome, pyrochlore, and other line-graph lattices  \cite{
Mielke1991FM, Mielke, Tasaki, Mielke2, Tasaki2, Mielke3, PhysRevB.78.125104, 
PhysRevLett.103.206805,  PhysRevLett.106.236803,  PhysRevLett.125.266403, Nakai:2022aa, BAB}. 
Frustration-freeness also allows one to derive rigorous results about the nature of the ground state(s) when electron-electron interactions are incorporated~\cite{Mielke1991FM, Mielke, Mielke2, Mielke3,Tasaki, Tasaki2, PhysRevB.92.115137,PhysRevB.98.155119,PhysRevB.105.024520}, as exemplified by the saturated ferromagnetism in partially filled flat bands~\cite{TasakiBook}. 
These results are particularly relevant in recent years due to the emergence of tunable flat-band materials in both natural~\cite{Ye2018, Liu2018, Yin2018, Kang2020, PhysRevB.102.075148, PhysRevMaterials.5.044202, Sales2022, Yin2022, PhysRevLett.128.096601, PhysRevResearch.5.043269, chen2024cascade, Wakefield2023, Huang2024}
and artificial~\cite{PhysRevLett.99.070401, PhysRevLett.108.045305, PhysRevLett.110.185302, PhysRevLett.110.185301,  Cao:2018aa, Cao:2018ab}
platforms.
Existing general results on frustration-free Hamiltonians typically concern bosonic systems like quantum magnets~\cite{Knabe,Gossetozgunov,Anshu,Lemm_2022,arXiv:2310.16881,lemm2024criticalfinitesizegapscaling,arXiv:2406.06415,arXiv:2406.06414}.
Motivated by the recently highlighted connections between frustration-freeness,  flat bands, and correlated topological materials~\cite{TasakiBook, PhysRevB.102.235101, BAB2}, we develop a general theory for frustration-free free fermions (F$^4$) in this work.

{\it F$^4$ conditions}---We begin by deriving the F$^4$ conditions for particle-number conserving Hamiltonians. 
Consider a free-fermion Hamiltonian $\hat H$ defined on a lattice
\begin{equation}
\hat{H}=\sum_{\bm{R}}\hat{H}_{\bm{R}}; \quad
\hat{H}_{\bm{R}}=\hat{\bm{c}}_{\bm{R}}^\dagger h_{\bm {R}}\hat{\bm{c}}_{\bm{R}}+C_{\bm{R}}.
\end{equation}
Here, $\hat H_{\bm R}$ is a local term defined around the point $\bm{R}$, 
$\hat{\bm{c}}_{\bm{R}}$ is an $N_r$-dimensional column vector with components given by $\hat{c}_{\bm{r} \sigma}$ within a range $|\bm r - \bm{R}| <r$, and 
$\sigma$ labels the fermion modes (e.g., sublattices, orbitals, and spin) assigned to $\bm r$. 
We assume that $\hat H$ is local in the sense that $r$ is finite and independent of the system size. The fermion hopping amplitudes are encoded in the $N_r$-dimensional Hermitian matrices $h_{\bm R}$.
The constant $C_{\bm R}$ will be fixed later.

Let $\mu_{\bm{R} \alpha}>0$ ($\alpha=1,2,\cdots,A_{\bm{R}}$) be the positive eigenvalues of $h_{\bm{R} }$ and $\bm{\psi}_{\bm{R}  \alpha}$ be the corresponding normalized eigenvectors. Likewise, let $-\nu_{\bm{R} \beta}<0$ ($\beta=1,2,\cdots,B_{\bm{R}}$) and $\bm{\phi}_{\bm{R} \beta}$ be the negative eigenvalues and associated eigenvectors.
We do not use the null vectors in our construction. Define
\begin{align}
\hat{ \psi}_{\bm{R}\alpha}\coloneqq \bm{\psi}_{\bm{R} \alpha}^\dagger\hat{\bm{c}}_{\bm{R}},\quad
\hat{\phi}_{\bm{R}\beta}\coloneqq \bm{\phi}_{\bm{R} \beta}^\dagger\hat{\bm{c}}_{\bm{R}}.\label{psiphi}
\end{align}
Operators with the same label $\bm{R}$ automatically satisfy canonical anticommutation relations.
Also, for any $\bm{R},\bm{R}'$
\begin{align}
&\{\hat{\psi}_{\bm{R}\alpha},\hat{\psi}_{\bm{R}'\alpha'}\}=\{\hat{\phi}_{\bm{R}\beta},\hat{\phi}_{\bm{R}'\beta'}\}=\{\hat{\psi}_{\bm{R}\alpha},\hat{\phi}_{\bm{R}'\beta}\}=0\label{AC3}.
\end{align}
For $\bm{R}\neq \bm{R}'$,  however, $\{\hat{\psi}_{\bm{R}\alpha},\hat{\psi}_{\bm{R}'\alpha'}^\dagger\}$, $\{\hat{\phi}_{\bm{R}\beta},\hat{\phi}_{\bm{R}'\beta'}^\dagger\}$, and $\{\hat{\psi}_{\bm{R}\alpha},\hat{\phi}_{\bm{R}'\beta}^\dagger\}$ do not vanish generally.

Setting $C_{\bm{R}}=\sum_{\beta}\nu_{\bm{R}\beta}$, the local Hamiltonian can be rewritten as $\hat{H}_{\bm{R}}=\hat{H}_{\bm{R}}^{(+)}+\hat{H}_{\bm{R}}^{(-)}$ with
\begin{align}
\hat{H}_{\bm{R}}^{(+)}\coloneqq\sum_{\alpha=1}^{A_{\bm R}}\mu_{\bm{R}\alpha}\hat{\psi}_{\bm{R}\alpha}^\dagger\hat{\psi}_{\bm{R}\alpha},\quad
\hat{H}_{\bm{R}}^{(-)}\coloneqq\sum_{\beta=1}^{B_{\bm R}}\nu_{\bm{R}\beta}\hat{\phi}_{\bm{R}\beta}\hat{\phi}_{\bm{R}\beta}^\dagger.\label{Hamiltonianpm}
\end{align}
Note that both $\hat{\psi}_{\bm{R}\alpha}^\dagger\hat{\psi}_{\bm{R}\alpha}$ and $\hat{\phi}_{\bm{R}\beta}\hat{\phi}_{\bm{R}\beta}^\dagger=1-\hat{\phi}_{\bm{R}\beta}^\dagger\hat{\phi}_{\bm{R}\beta}$ are projectors with eigenvalues $0$ and $1$. Therefore, $\hat{H}_{\bm{R}}$ is positive semi-definite. 

Our discussion up to this point applies generally to any number-conserving free-fermion problems.
We now introduce our first result---the necessary and sufficient condition for F$^4$ models. Recall that a local Hamiltonian $\hat H$ is frustration-free if it admits a decomposition $\hat{H}=\sum_{\bm{R}}\hat{H}_{\bm{R}}$ with positive-semidefinite local terms $\hat{H}_{\bm{R}}$ and a state $|\Phi\rangle$ such that $\hat{H}_{\bm{R}}|\Phi\rangle=0$ for all $\bm{R}$ simultaneously~\cite{TasakiBook}.
Given the form in Eq.\ \eqref{Hamiltonianpm}, we claim
\begin{align}
\hat H ~\text{is F$^4$} 
\quad \Longleftrightarrow \quad
\{\hat{\psi}_{\bm{R}\alpha},\hat{\phi}_{\bm{R}'\beta}^\dagger\}=0\quad \forall \bm{R} , \bm{R}', \alpha, \beta.\label{conditiontobeFFFF}
\end{align}
A sketch of the proof is provided below; a more detailed discussion can be found in Ref.~\cite{masaoka-ono-po-watanabe_FF25}.
If $\hat H$ is F$^4$, then its ground state $|\Phi \rangle$ satisfies $\hat{\psi}_{\bm{R}\alpha} | \Phi \rangle = \hat{\phi}_{\bm{R}'\beta}^\dagger|\Phi\rangle = 0$ for all $\bm{R}, \alpha, \beta$. As $\{\hat{\psi}_{\bm{R}\alpha},\hat{\phi}_{\bm{R}'\beta}^\dagger\}$ is a $\mathbb{C}$-number, the existence of $|\Phi\rangle$ implies the anti-commutator vanishes. In the other direction, one can define a state $| \Phi \rangle$ by applying all of the linearly independent fermion creation operators among the set $\{ \hat \phi_{\bm R, \beta }^\dagger \}_{\bm R, \beta}$ to the vacuum. The vanishing of the anti-commutator in Eq.\ \eqref{conditiontobeFFFF} then implies $\hat H_{\bm R}|\Phi \rangle =0$ for all $\bm{R}$.

We note that the condition Eq.\ \eqref{conditiontobeFFFF} also generalizes naturally to fermionic bilinear Hamiltonians without particle U(1) symmetry, which, as we discuss in Ref.\ \cite{masaoka-ono-po-watanabe_FF25}, goes beyond some existing results~\cite{PhysRevResearch.3.033265, jones2023exact}.

{\it Excitation gap of gapless models}---As a first application of the F$^4$ condition, we prove that the excitation gap $\epsilon(L)$ scales as $O(1/L^2)$ for gapless F$^4$ Hamiltonians with particle U(1) and lattice translation symmetries, where $L$ denotes the finite linear size of the system. 

More concretely, consider a F$^4$ Hamiltonian $\hat H(L)$ with a variable finite linear size $L$. Let $\epsilon(L)>0$ be the first non-zero eigenvalue of $\hat H(L)$ (possibly in a different fermion number sector). We say the system is gapless if $\epsilon(L)$ approaches $0$, the ground-state energy, in the thermodynamic limit. 
From the decomposition in Eq.\ \eqref{Hamiltonianpm}, we can write the full Hamiltonian as $\hat H = \hat H^{(+)} + \hat H^{(-)}$, where $\hat H^{(\pm)} = \sum_{\bm R} \hat H^{(\pm)}_{\bm R}$. By construction, the non-zero single-particle energies $\omega^{(+)}_{\gamma}$ associated with $\hat H^{(+)}$ (resp. $\omega^{(-)}_{\gamma}$ for $\hat H^{(-)}$) are all positive (negative). 
Furthermore, the F$^4$ condition Eq.\ \eqref{conditiontobeFFFF} implies $[\hat{H}_{\bm{R}}^{(+)},\hat{H}_{\bm{R}'}^{(-)}] = 0$ and therefore the non-zero single-particle energies of $\hat H$ can be obtained by combining those from $\hat H^{(\pm)}$. Upon Fourier transform,  the Bloch Hamiltonians associated with $\hat H^{(+)}$ and $\hat H^{(-)}$ commute at every momentum $\bm{k}$ in the Brillouin zone (BZ). 
The mentioned non-zero single-particle energies of $\hat H$ can then be labeled by the momentum $\bm k$ as $\omega^{(\pm)}_{\bm k, n}$, where $n=1,2,\dots$ is counted from zero energy. 

As $\omega^{(\pm)}_{\bm k, n}$ are independently defined from $\hat H^{(\pm)}$, one heuristically expects any band touching between the two sets, if any, must be quadratic or softer. The finite-size sampling on the dispersion curve then leads to the asserted finite-size gap scaling.
This heuristic can be justified more carefully through the construction of an auxiliary Dirac Hamiltonian.
For concreteness, we focus on $\hat H^{(+)}$ below; the construction for $\hat H^{(-)} $ is similar.
Starting from Eq.\ \eqref{Hamiltonianpm}, we can construct an auxiliary Hamiltonian $\hat {\tilde H}^{(+)} = \sum_{\bm{R}} \hat {\tilde H}_{\bm R}^{(+)}$, where
\begin{equation}\begin{split}\label{eq:auxiliaryhamiltonian}
\hat{\tilde H}_{\bm R}^{(+)} = 
\left(
\begin{array}{cc}
\hat {\bm{c}}_{\bm{R}}^\dagger  & 
\hat {\bm{d}}_{\bm{R}}^\dagger
\end{array}
\right)
\left(
\begin{array}{cc}
0 &  {\psi}_{\bm {R}} \sqrt{\mu_{\bm R}}\\
 \sqrt{\mu_{\bm R}} {\psi}_{\bm {R}}^\dagger & 0
\end{array}
\right)
\left(
\begin{array}{c}
\hat {\bm{c}}_{\bm{R}}\\
\hat {\bm{d}}_{\bm{R}}
\end{array}
\right).
\end{split}\end{equation}
Here, $\hat {\bm {d}}_{\bm{R}} = (\hat d_{\bm{R}1}, \hat d_{\bm{R}2}, \dots, \hat d_{\bm{R}A_{\bm{R}} })^T$ is a vector of auxiliary fermion annihilation operators introduced at $\bm{R}$, $\mu_{\bm{R}}$ is a diagonal matrix consisting of the original local eigenvalues $\mu_{\bm{R} \alpha} >0$ for $\alpha = 1,\dots, A_{\bm{R}}$, and $\psi_{\bm{R}}$ is an $N_r\times A_{\bm{R}}$ dimensional matrix formed by the eigenvectors $\bm{\psi}_{\bm{R}\alpha}$ of $h_{\bm {R}}$. 

As defined, $\hat {\tilde H}^{(+)}$ is a local Hamiltonian with chiral symmetry, and so its non-zero single-particle energies come in pairs of the form $\pm \nu^{(+)}_\gamma$. 
Although the ``square root'' is taken for each local term $\hat H_{\bm R}^{(+)}$, one can verify that the decoupled block in $\left( \tilde h^{(+)} \right)^2$ acting on the $\hat c$-fermions is identical to the single-particle Hamiltonian $h^{(+)}$ of the original problem (Appendix \ref{app:dirac}). This establishes $\omega^{(+)} = \left( \nu^{(+)}_{\gamma}\right)^2$. 
Furthermore, the auxiliary Dirac-like Hamiltonians are also translation-invariant, and we have the momentum-resolved relation $\omega^{(\pm)}_{\bm k, n} = \pm \left( \nu_{\bm k, n}^{(\pm)}\right)^2$. 
This reveals that the non-zero bands of any F$^4$ model can always be understood as the square of some auxiliary bands.

If $\hat H$ is gapless, at least one of $\nu_{\bm k, 1}^{(+)}$ and $\nu_{\bm k, 1}^{(-)}$ must vanish at some momentum $\bm k_*$. Consider a small line segment $\bm k = \bm k_* + \delta k \hat {\bm \kappa}$ passing through $\bm k_*$ for any direction $\hat {\bm \kappa}$.
As the auxiliary Hamiltonians only contain finite-range hopping, with suitable labeling the dispersion $\nu^{(\pm)}_{\bm k, n}$ can also be expanded analytically in $\delta k$~\cite{rellich1969perturbation}. Now, if $\nu_{\bm k_*, 1}^{(+)} = 0$, we have $\nu^{(+)}_{\bm {k}, 1} = v_{\hat{\bm \kappa}} \delta k +  O(\delta k^2)$ for some velocity component $v_{\hat{\bm \kappa}}$.
We can then conclude a quadratic dispersion $\omega_{\bm{k}1}^{(+)} = v_{\hat{\bm \kappa}}^2 \delta k^2 + \mathcal  O(\delta k^3)$ around $\bm{k}= \bm{k}_* + \delta k \hat {\bm{\kappa}} $.
The $O(1/L^2)$ finite-size scaling of excitation energy follows from the discrete sampling of the BZ.

\begin{figure}[t]
\begin{center}
{\includegraphics[width=0.48 \textwidth]{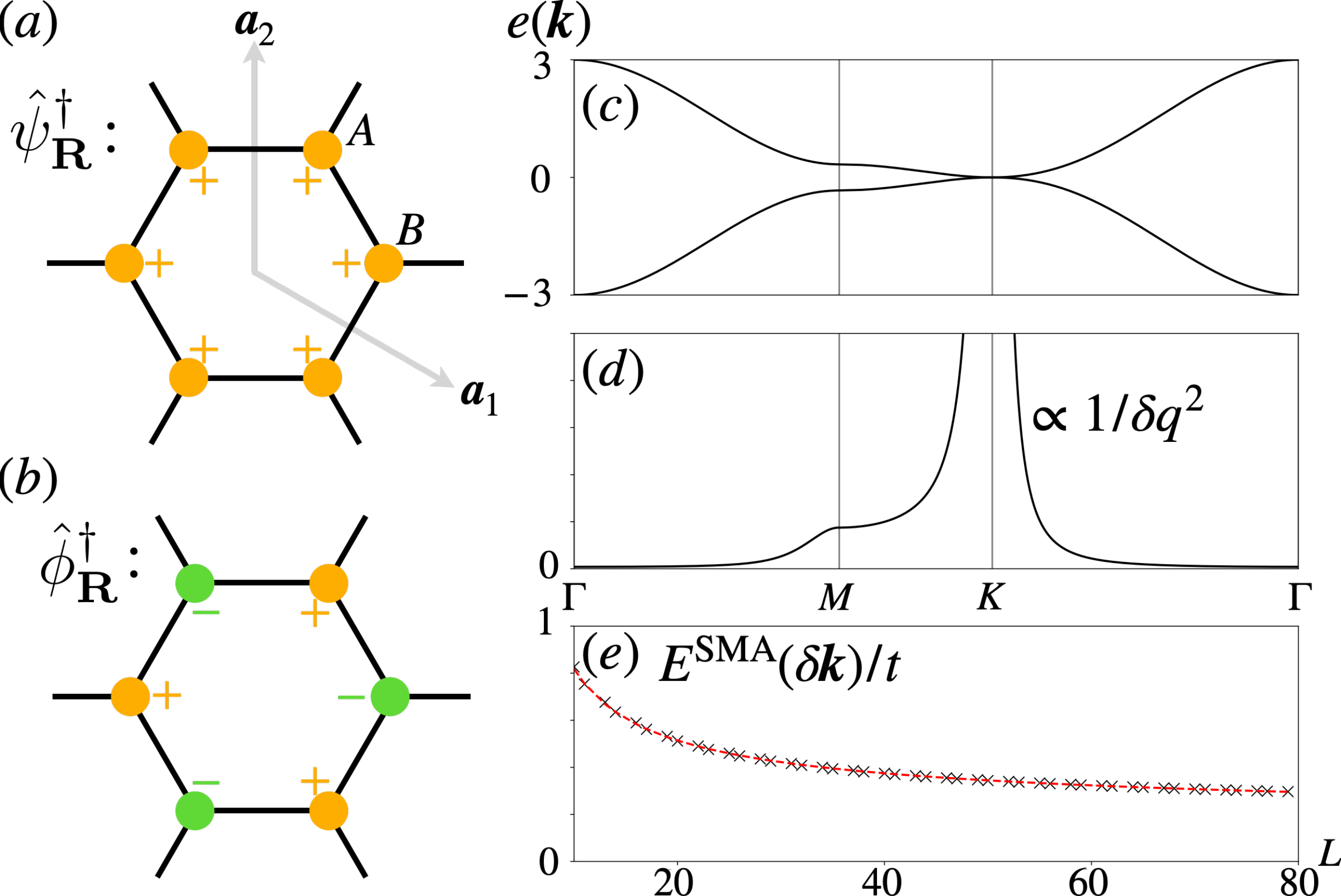}} 
\caption{ 
Frustration-free free-fermion model on the honeycomb. (a) Lattice convention and the definition of the local orbital corresponding to the conduction band. (b) Local orbital for the valence band. (c) Band structure. (d) Non-zero eigenvalue of the quantum metric tensor in arbitrary units, which diverges as $1/\delta q^2$ around the K point. (e) Excitation energy of density fluctuations within the single-mode approximation, which acquires an anomalous gap $O(1/\log L)$ due to the singular quantum metric. We take $\delta \bm k = (\bm b_1 + \bm b_2)/L$. Crosses are computed values and the dashed line indicates an inverse-log fit with fitting parameters reported in Appendix \ref{app:honeycomb_details}.
\label{fig:orbitals_sma}
 }
\end{center}
\end{figure}

{\it Honeycomb model}---We now illustrate our results through an example. Consider spinless fermions hopping on the two-dimensional honeycomb lattice with one $s$ orbital per site  (Fig.\ \ref{fig:orbitals_sma}). 
There are symmetry-protected band touchings at the K and K' points. Although linear dispersion, like that is graphene, is generally expected, our results imply that any F$^4$ Hamiltonians must have the Dirac fermions softened into band touching points that are quadratic or softer, i.e., the F$^4$ requirement automatically tunes the Dirac velocity to zero.

Let $\hat c^\dagger_{{\bm R}, \sigma}$ for $\sigma = A,B$ be spinless fermions defined on the $\sigma$ site of unit cell ${\bm R}$.
We consider a finite $L$-by-$L$ lattice along the two lattice vectors.
Based on the F$^4$ condition Eq.\ \eqref{conditiontobeFFFF}, we define compactly supported fermions $\hat \psi_{\bm {R}}$ and $\hat \phi_{\bm{ R} }$ by taking a superposition of the six orbitals around a hexagon, as shown in Fig.\ \ref{fig:orbitals_sma} (a,b). One can verify $ \{\hat{\psi}_{\bm{R}},\hat{\phi}_{\bm{R}'}^\dagger\}=0$ and therefore 
$\hat H = t\sum_{\bm{R}} \left( \hat \psi_{\bm {R}}^\dagger \hat \psi_{\bm {R}} + 
 \hat \phi_{\bm {R}} \hat \phi_{\bm {R}}^\dagger 
\right)$
is a F$^{4}$ Hamiltonian. 

Introduce Fourier transform with $\hat c_{{\bm k}, \sigma}^\dagger = \frac{1}{L} \sum_{{\bm R} } \hat c_{{\bm R}, \sigma}^\dagger e^{i {\bm k} \cdot {\bm R} }$:
\begin{equation}\begin{split}\label{eq:}
\hat \psi_{{\bm k} }^\dagger = \hat {\bm c}^\dagger_{{\bm k} } {\bm \psi}_{{\bm k}};\quad
{\bm \psi}_{{\bm k}} =
\frac{1}{\sqrt{6}} 
\left(
\begin{array}{c}
1 + e^{i 2 \pi k_2}  + e^{i 2 \pi (k_1+k_2)}\\
1 + e^{i 2 \pi k_1}  + e^{i 2 \pi (k_1+k_2)}
\end{array}
\right)
\end{split}\end{equation}
where we write ${\bm k} = k_1 {\bm b}_1 + k_2 {\bm b}_2$ with ${\bm b}_{1,2}$ denoting the reciprocal lattice vectors and  $\hat {\bm c}^\dagger_{{\bm k} }  = (\hat c_{{\bm k}, A}^\dagger , \hat c_{{\bm k}, b}^\dagger )$. $\hat \phi_{{\bm k}}^\dagger$ is similarly defined but with ${\bm \phi}_{{\bm k}} = \sigma_3 {\bm \psi}_{{\bm k}}$. Here, $\sigma_3$ is the Pauli matrix in the sublattice space. 
Since the fermions defined on neighboring unit cells do not satisfy canonical anti-commutation relations,  ${\bm \psi}_{{\bm k}} $ and ${\bm \phi}_{{\bm k}} $ as defined are not normalized. 
Instead, one has $|{\bm \psi}_{{\bm k}}|^2 = |{\bm \phi}_{{\bm k}}|^2 = e_{\bm k}$, where
\begin{equation}\begin{split}\label{eq:}
e_{\bm k} = 1 + \frac{2}{3} \left( 
\cos(2 \pi k_1)+\cos(2 \pi k_2)+\cos(2 \pi (k_1+k_2))
\right).
\end{split}\end{equation}
Notice that $e_{\bm k} =0$ at the K and K' points ${\bm k} = (\pm 1/3, \pm 1/3)$ and so ${\bm \psi}_{{\bm k}} $ and ${\bm \phi}_{{\bm k}} $ vanish there.
This implies that the sets of localized orbitals $\{ \hat \psi^\dagger_{\bm R}\}$ and  $\{ \hat \phi^\dagger_{\bm R}\}$ are not linearly independent over the full lattice.

In momentum space, ${\hat H = \sum_{\bm k} \hat {\bm c}^\dagger_{{\bm k} } h({\bm k}) \hat {\bm c}_{{\bm k} }} + L^2 $ with the Bloch Hamiltonian 
$h({\bm k}) = {t ( {\bm \psi}_{{\bm k}}{\bm \psi}_{{\bm k}}^\dagger - {\bm \phi}_{{\bm k}} {\bm \phi}_{{\bm k}}^\dagger)}$. 
As condition Eq.\ \eqref{conditiontobeFFFF} implies ${\bm \psi}_{{\bm k}}^\dagger {\bm \phi}_{{\bm k}}=0$, 
for $e_{\bm k} \neq 0$ the normalized eigenvectors for the conduction and valence bands are respectively $\bm{u}_{\bm{k}, c} = {\bm \psi}_{{\bm k}}/\sqrt{e_{\bm k}}$ and $\bm{u}_{\bm{k}, v} = {\bm \phi}_{{\bm k}}/\sqrt{e_{\bm k}}$. The associated fermion creation operators are $\hat f^\dagger_{\bm k, \alpha} = \hat {\bm c}_{\bm k}^\dagger \bm{u}_{\bm{k}, \alpha}$ for $\alpha = c,v$.
The band dispersions $\pm t e_{\bm k}$ touch quadratically at $\bm k =$ K and K' (Fig.\ \ref{fig:orbitals_sma}(c)), as required by the F$^4$ condition. 
Furthermore, taking the limit $\bm{k} \to \text{K}, \text{K}'$ reveals that the normalized Bloch states $\bm{u}_{\bm{k}, \alpha}$ are ill-defined at these gap closing points, indicating that the quadratic band touching is singular~\cite{PhysRevB.99.045107} .

Next, we turn to the finite-size excitation gap.
The quadratic band touching immediately indicates that the single-particle excitation gap goes as $O(1/L^2)$. Yet, it is also natural to ask if the charge-neutral excitation gap vanishes in the same way. 
The answer is again positive, as we can consider the particle-hole excitations in the band basis
\begin{equation}\begin{split}\label{eq:ph_trial}
|\bm k, \bm q\rangle = \hat f^\dagger_{\bm q + \bm k, c} \hat f_{\bm q, v} |\Phi \rangle,
\end{split}\end{equation}
where $ | \Phi \rangle$ is the ground state, $\bm k$ labels the many-body momentum of the ansatz state, and the extra momentum $\bm q$ should be viewed as a label indicating the momentum of the created hole. 
Evidently, $|\bm k, \bm q\rangle $ is an energy eigenstate of $\hat H$ with eigenvalue $t (e_{\bm k + \bm q} + e_{\bm q})$, and it disperses quadratically with $\bm k$ when $\bm q$ approaches K and K'. This shows that the charge-neutral excitation gap also vanishes as $O(1/L^2)$.

{\it Interaction and density fluctuations}---
Does the same excitation gap scaling hold beyond the non-interacting limit? For a general fermionic frustration-free model, we have shown in Ref.\ \cite{masaoka-ono-po-watanabe_FF25} that,
assuming a power-law decay of a two-point correlation function in the ground state, the Gosset-Huang inequality~\cite{GossetHuang} implies $\epsilon(L) = O((\log L)^2 /L^2)$. 
We now turn to validating this assertion when a local repulsion $
\hat U = U \sum_{\bm R}  (\hat \psi^\dagger_{\bm R}\hat \psi_{\bm R}) (\hat \phi^\dagger_{\bm R}\hat \phi_{\bm R})$
with $U>0$ is added to the F$^4$ model. By design, each local term in $\hat U$ is positive semi-definite and annihilates the non-interacting ground state $|\Phi \rangle$, and so $\hat H^{\rm int} = \hat H + \hat U$ remains a frustration-free Hamiltonian. 
Yet, $\hat U$  modifies the excitation energies and so the behavior of $\epsilon(L)$ should be reexamined. 

To bound $\epsilon(L)$, it suffices to consider suitable variational excited states. 
In the context of interacting frustration-free bosonic systems, this is often achieved through the density fluctuations treated within the single-mode approximation (SMA)~\cite{SMA}. The trial states take the form
$| {\bm k} \rangle \propto \sum_{{\bm R}} \hat n_{\bm R} e^{i {\bm  k} \cdot {\bm R} } | \Phi \rangle$, where  $\hat n_{\bm R} $ is the total density operator in the unit cell ${\bm R}$. 
The SMA energy is given by the energy expectation value $E^{\rm SMA}({\bm k}) = \langle {\bm k} | \hat H | {\bm k} \rangle/{\langle {\bm k} | {\bm k} \rangle }$ and satisfies $E^{\rm SMA}({\bm k}) = O(|{\bm k}|^2)$ in many known models~\cite{PhysRevLett.131.220403, arXiv:2405.00785, han2025modelsinteractingbosonsexact, arXiv:2406.06414}.

Surprisingly, this commonly used ansatz does not provide a useful bound in the present honeycomb model. Instead, due to a diverging quantum metric in the vicinity of the gap closing point of the non-interacting limit, $E^{\rm SMA}$ acquires an anomalous gap of $O(1/\log L)$ \cite{masaoka-ono-po-watanabe_FF25}.
Nevertheless, we can still demonstrate $\epsilon(L) = O(1/L^2)$ using variational excited states in Eq.\ \eqref{eq:ph_trial}. Consider zero-momentum particle-hole excitations $| \bm 0, \bm q\rangle$. When $U=0$, the excitation energy for $\bm q$ around the K and K' points takes the form $ 2t e_{\bm q_L}^{\rm min} $ with $ e_{\bm q_L}^{\rm min} = \min \{ e_{\bm q} > 0 \} = O(1/L^2)$.
As shown in Appendix \ref{app:honeycomb_details}, 
a similar bound of $(2t+U) e_{\bm q_L}^{\rm min} $ can be derived when $U>0$.
Analyzing $e_{\bm q_L}^{\rm min} $ further, we obtain
\begin{equation}\begin{split}\label{eq:zero_momentum_ph}
\epsilon(L)\leq  
(2t+U) 2 \pi^2 \frac{1}{L^2} + O(1/L^3)
\end{split}\end{equation}
which provides the desired $O(1/L^2)$ bound. 

\begin{figure}[t]
\begin{center}
{\includegraphics[width=0.48 \textwidth]{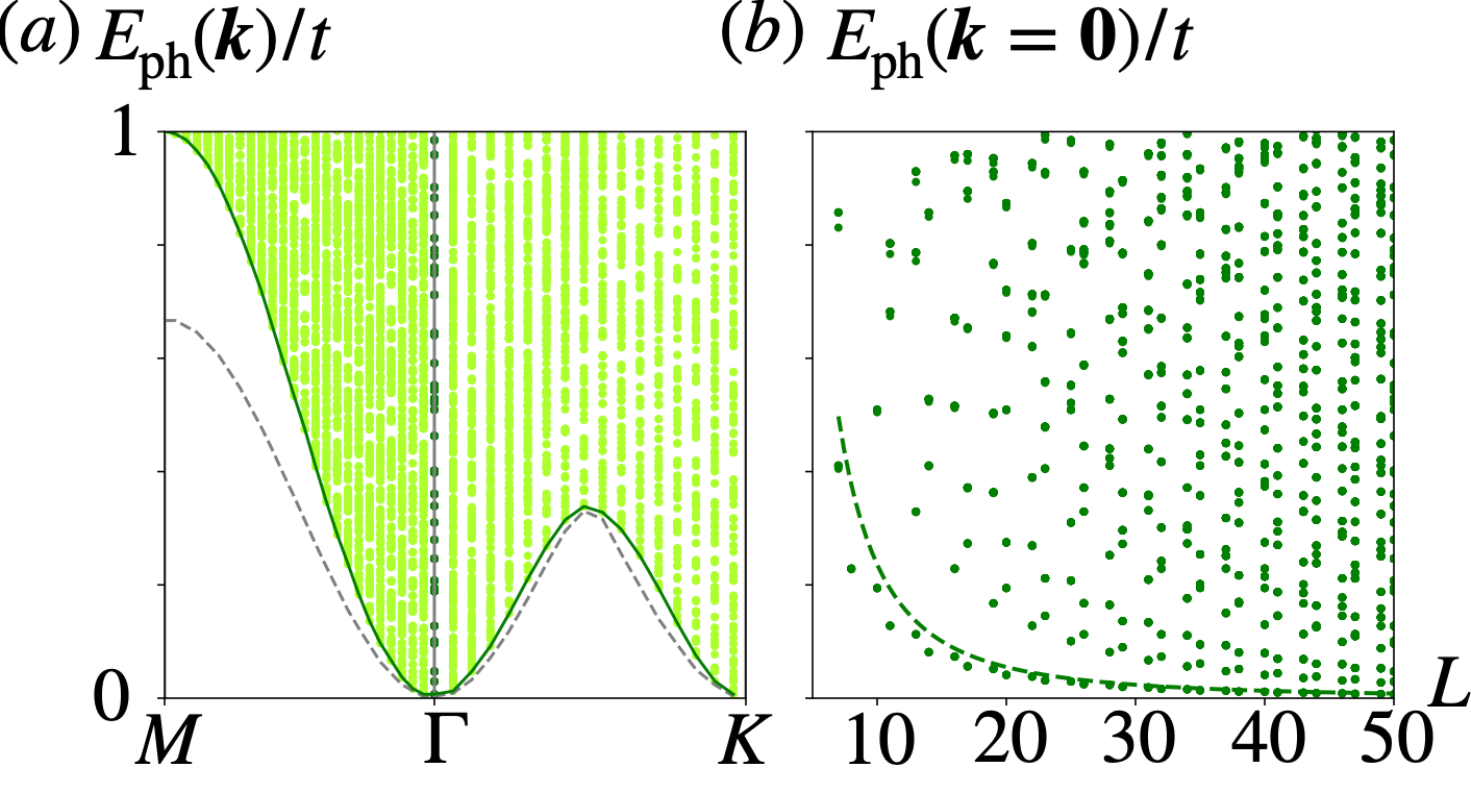}} 
\caption{Energy of particle-hole excitations in $\hat H^{\rm int}$ at $U=2t$. (a) Energy spectrum of $\hat H^{\rm int}$ restricted to the subspace spanned by $|\bm k, \bm q\rangle$ for $\bm k$ along the M-$\Gamma$-K line computed for $L= 50$. 
The boundary of the particle-hole continuum can be estimated as $\min_{\bm q} (t e_{\bm q} + (t+U) e_{\bm q + \bm k}) $ (green solid line), which becomes exact in the non-interacting limit with $U=0$ (grey dashed line). Note that interaction $U>0$ modifies the excitation energies although it keeps the ground states unchanged. 
(b) System size dependence of the particle-hole energy at $\bm k = \bm 0$. The dashed line shows the $O(1/L^2)$ bound in Eq.\ \eqref{eq:zero_momentum_ph}.
\label{fig:ph_energy}
 }
\end{center}
\end{figure}

In Fig.\ \ref{fig:ph_energy}, we show that the particle-hole excitation indeed disperses quadratically around $\Gamma$, confirming the discussed finite-size bound on the minimum excitation energy. Numerically, the corresponding bound of the charge-neutral finite-size gap decays as $L^{-1.99(2)}$ (Appendix~\ref{app:honeycomb_details}), consistent with our analytical results.

{\it Correlation functions}---
Having derived the general condition and illustrated it through a 2D honeycomb model, we next demonstrate how the F$^4$ conditions inform the general behavior of ground state correlation functions.

Generally, the ground state correlation functions of a gapped, local Hamiltonian exhibit exponential decay~\cite{PhysRevLett.93.140402,hastingsSpectralGapExponential2006}. Conversely, while a power-law decay is naturally anticipated in the ground state of a gapless system, there can also be gapless models in which all ground state correlation functions decay exponentially with distance. 
Both possibilities are possible in frustration-free models.
For instance, the uncle Hamiltonian~~\cite{uncle,arXiv:2406.06414} for matrix product states is gapless but its ground state has only exponentially decaying correlation functions. In contrast, the ground state of the Motzkin chain~\cite{PhysRevLett.109.207202,menon2024symmetriescorrelationfunctionsentanglement} exhibits power-law correlation functions with critical exponents $z\geq 2$.  

We now investigate how the F$^4$ condition informs the behavior of the ground-state correlations.
A free-fermion model is gapless if and only if there is a fermion mode with energy $\epsilon=0$ in the thermodynamic limit. 
While a zero-energy mode implies ground state degeneracy, one can ask: does there exist one ground state exhibiting only exponentially decaying correlation functions? We say the gapless model is non-critical when such a ground state exists.

A free-fermion model must be non-critical if there is a continuous band gap separating the conduction and the valence bands in the thermodynamic limit. It remains to consider the case when there is a band touching point at $\epsilon =0$.
From our proof, we know that a ground state $|\Phi\rangle$ of any F$^4$ Hamiltonian can be constructed by occupying all the locally defined $\phi$ orbitals. 
Let $\epsilon_{v}^*$ denote the top of the valence bands, defined as the maximum single-particle energy among the valence bands over the continuous BZ.
When $\epsilon_{v}^* =0$, the $\phi$ orbitals are incomplete in the following sense:
Suppose there are $N_v$ valence bands. 
Let $V$ be the system volume. From the general discussion on the excitation gap, we see that $\epsilon_{v}^* =0$ if and only if there are fewer than $V N_v$ linearly independent orbitals among the set $\{ \hat \phi_{\bm R, \beta}\}_{\bm R, \beta}$. 

Importantly, incompleteness of the $\phi$ orbitals is not equivalent to criticality:
If there exist additional locally defined orbitals $\hat \phi'_{\bm R' \beta'}$ satisfying $\{ \hat \phi'_{\bm R' \beta'}, \hat \psi_{\bm R\alpha}\} = 0$, they can be added to the original Hamiltonian without altering the conduction bands. 
When such $\phi'$ orbitals can be found such that the combined set of $\phi$ and $\phi'$ orbitals contains $V N_v$ linearly independent operators, the modified F$^4$ Hamiltonian yields $\epsilon_{v}^* <0$. 
Since the ground state of the modified, gapped Hamiltonian is also a ground state of the original gapless model, we conclude the original model is non-critical.
The desired $\phi'$ orbitals may not exist.
For instance, in the case of the nearest-neighbor kagome model, the set of flat-band orbitals can only be completed by including extended states that are delocalized along at least one direction of the system~\cite{PhysRevB.78.125104}. 

Naturally, the feasibility to complete the set of $\phi$ orbitals is tied to the analytical properties of the Bloch wave functions around the gap closing point. 
Recall, our F$^4$ condition implies any gapless F$^4$ model can be deformed to a flat-valence-band limit by setting $\nu_{{\bf R}\beta }\rightarrow 0$. This allows us to analyze the criticality problem in the framework of flat-band singular band touching
\cite{PhysRevB.99.045107}. When there is a singular band touching point, the Bloch states in its vicinity must be non-analytic the ground state correlation functions exhibit power-law decay \cite{masaoka-ono-po-watanabe_FF25}.
Conversely, for a single-valence-band problem with only non-singular band touching points, Ref.\ \cite{PhysRevB.99.045107} showed that a band gap can open without modifying the Bloch states. Such scenario corresponds to non-critical gapless F$^4$ model, which is generally the case for one-dimensional models~\cite{masaoka-ono-po-watanabe_FF25}. 

The analysis above motivates the following conjecture: a gapless F$^4$ model is non-critical if and only if, in the flat-band limit, all the band touchings between the conduction and valence bands are non-singular. While the ``only if'' direction has been established~\cite{masaoka-ono-po-watanabe_FF25}, the ``if'' part is currently known only for the special case of a single valence band~\cite{PhysRevB.99.045107}.

{\it Gapped topological bands}---
We now turn to gapped F$^4$ models.
It is well-known that the Kitaev model for a 1D topological superconductor admits a F$^4$ realization. 
What other forms of band topology are compatible with the F$^4$ condition?
For concreteness, we assume there is a band gap and the bottom of the conduction bands, defined in the thermodynamic limit, satisfies $\epsilon_c^*>0$.
We now show that the only possible tenfold-way topological phases realizable in the valence bands would be weak phases obtained by stack of 1D phases.
We further conjecture that any other stable crystalline topological phases are similarly forbidden.
In addition, we show that fragile topology is generally compatible with the F$^4$ conditions.

We argue by first addressing the possible topology of the conduction bands.
Given the gap condition $\epsilon_c^*>0$, the $\psi$ orbitals in the defining F$^4$ Hamiltonian provide a set of compact orbitals which span all the states in the conduction band. 
Note that these orbitals might be over-complete, which happens when the number of $\psi$ orbitals in each unit cell exceeds the number of conduction bands. Nevertheless, Ref.~\cite{PhysRevB.95.115309} proved that the only ten-fold way topology realizable in such bands are weak phases obtained by stacking 1D topological phases. 
As the stable topological indices for the whole system is necessarily zero, we conclude the only possible ten-fold way topology of the valence bands will also be weak phases stemming from the 1D classification. 
Although not addressed explicitly in Ref.~\cite{PhysRevB.95.115309}, the analysis therein could be enriched by additional crystalline symmetries, and so we conjecture that other forms of stable topology beyond those deformable to representatives built from 0D and 1D topological building blocks~\cite{TC_PRB, TC_PRX, R-AHSS_Song, R-AHSS_Shiozaki,defect_network, Shiozaki-Ono2023, Ono-Shiozaki-Watanabe2022, WC_Fang} are similarly forbidden in gapped F$^4$ Hamiltonians with a gap $\epsilon_c^*>0$.

Fragile topology~\cite{PhysRevLett.121.126402}, which arises frequently as the complement of atomic bands, could still exist even when stable topology is ruled out. Indeed, when one takes the flat-band limit for the fragile topological model in Ref.~\cite{PhysRevB.99.195455} for twisted bilayer graphene, the resulting Hamiltonian satisfies our F$^4$ condition. In Ref.~\cite{masaoka-ono-po-watanabe_FF25}, we further demonstrate that the valence bands could acquire dispersion without breaking the F$^4$ conditions.

{\it Conclusions and outlook}---In this work, we derived the necessary and sufficient conditions for local frustration-free free-fermion systems, and proved that the touching between conduction and valence bands in such models must be quadratic or softer.
We also provided a concrete model on the honeycomb lattice, and found that the diverging quantum metric led to an anomalous gap in the density fluctuations. Nevertheless, one could construct charge-neutral excitations which bound the finite-size scaling of the excitation gap to be $O(1/L^2)$ even in the presence of interactions, consistent with the general results in Ref.\ \cite{masaoka-ono-po-watanabe_FF25}. We further show that gapless F$^4$ models can be non-critical, and that the criticality is tied to the presence of singular band touching points in the flat-band limit~\cite{PhysRevB.99.045107}. In addition, under a gap condition, we show that the valence bands of a F$^4$ model can only be atomic, weak-topological with 1D building blocks, or fragile topological.

Our results also lead to a collection of no-go concerning the possibility of realizing various topological phases using F$^4$ Hamiltonains. The dispersion bound rules phases like Dirac and Weyl semimetals with linear dispersion. Under the gap condition \footnote{
We remark that our conditions leave open one technical possibility: if a continuous band gap exists between the conduction and valence bands although $\epsilon_v^*=\epsilon_c^*=0$, and that both sets cannot be made complete by supplying symmetric, local orbitals, then our analysis is silent on the band topology which could be realized.
}, we also rule out the realization of any Wannier-obstructing tenfold-way topology~\cite{Po2017, Bradlyn2017, PhysRevB.97.035139}, such as Chern and $\mathbb{Z}_2$-topological insulators~\cite{PhysRevLett.98.046402, PhysRevB.83.035108,PhysRevB.92.205307, PhysRevB.95.115309}.
However, we remark that once interactions are included, frustration-free fermionic models can host nontrivial topology forbidden in the free-fermion limit. This has been demonstrated by the construction of commuting projector Hamiltonians for the $\mathbb{Z}_2$ topological insulator~\cite{PhysRevB.108.L121104} and recent developments in fermionic projected entangled pair states (PEPS)~\cite{PhysRevLett.132.126504,Xu_2024}.

\begin{acknowledgments}
We thank Hosho Katsura, Ken Shiozaki, Hal Tasaki, Carolyn Zhang, Ruben Verresen, and Shenghan Jiang for useful discussions.
H.C.P.~acknowledges support from the National Key R\&D Program of China (Grant No. 2021YFA1401500) and the Hong Kong Research Grants Council (C7037-22GF).
The work of H.W. is supported by JSPS KAKENHI Grant No.~JP24K00541.
S.O.~was supported by RIKEN Special Postdoctoral Researchers Program.
H.W., S.O., and H.C.P.~thank the Yukawa Institute for Theoretical Physics, Kyoto University, where we started the project during the YITP workshop YITP-T-24-03 on ``Recent Developments and Challenges in Topological Phases.''
\end{acknowledgments}

\bibliography{refs}

\clearpage
\appendix

\section{Details of the Dirac trick 
\label{app:dirac}}
Here, we explain why the Dirac trick applies to the full system even when the square root is apparently taken only for each of the local terms.

For concreteness, we focus on the original positive part of the F$^4$ Hamiltonian, given by $\hat H^{(+)} = \sum_{\bm R} \sum_{\alpha=1}^{A_{\bm R}} \mu_{\bm R\alpha} \hat \psi_{\bm R\alpha}^\dagger \hat \psi_{\bm R\alpha}$. Define a compound index $\eta$ which runs over all of $\bm R$ and $\alpha$. Also, let $\hat {\bm c}^\dagger$ be a row vector containing all the fermion creation operators in the system. We may write the locally defined fermions as $\hat \psi_{\eta}^\dagger = \hat {\bm c} \boldsymbol{ \psi}_{\eta}$. Here, $\boldsymbol{ \psi}_{\eta}$ is a column vector which is mostly zero, as the corresponding fermion is localized around the unit cell $\bm R$ associated with $\eta$. With this notation, one has $\hat H^{(+)} = \hat {\bm c}^\dagger  \left( \sum_{\eta} \boldsymbol{ \psi}_{\eta} \mu_\eta \boldsymbol{ \psi}_{\eta}^\dagger \right) \hat {\bm c}$. Let $Q$ be the matrix, generally not square, with the $\eta$-th column being $\boldsymbol{ \psi}_{\eta} \sqrt{\mu_\eta}$. The single-particle Hamiltonian of $\hat H^{(+)} $ is given by $h^{(+)} = Q Q^\dagger$. 
Noticing $\eta$ also indexes the auxiliary fermions introduced in the Dirac trick, one sees that
\begin{equation}\begin{split}\label{eq:}
\tilde h^{(+)} = \left(
\begin{array}{cc}
0 & Q\\
Q^\dagger & 0
\end{array}
\right)
\end{split}\end{equation}
when we combine all the local terms in defining the full-system single-particle matrix. Therefore, $(\tilde h^{(+)} )^2$ contains $h^{(+)}$ as a block.

\section{Details of the honeycomb model
\label{app:honeycomb_details}}
{\it Tight-binding form}---Consider a chiral symmetric honeycomb model of spinless fermions. Let the nearest-neighbor hopping be $t_1$. Assuming chiral symmetry represented by $\sigma_3$ in the sublattice space, the next allowed hopping $t_2$ is between third nearest neighbors. In momentum space, the Bloch Hamiltonian is given by $h_{\bm k} = Q_{\bm k} \sigma^+ + Q_{\bm k}^* \sigma^-$, where 
\begin{equation}\begin{split}\label{eq:}
Q_{\bm k} = &
t_1 \left (1 + e^{-i \bm k \cdot \bm a_1} + e^{i \bm k \cdot \bm a_2} \right) \\
&+ t_2 \left(e^{i \bm k \cdot (\bm a_1 +\bm a_2)} + e^{i \bm k \cdot (-\bm a_1 + \bm a_2)} + e^{- i \bm k \cdot (\bm a_1 + \bm a_2)} \right).
\end{split}\end{equation}
One can verify that the F$^4$ Hamiltonian defined in the main text gives the same model with $t_1 = \frac{2}{3}t $ and $t_2 = \frac{1}{3} t$. 
Conversely, the model is not F$^4$ for other values of $t_1/t_2$.

For general ratios of $t_1/t_2$, the K and K' points feature Dirac fermions. 
Expanding at the K point, we have
\begin{equation}\begin{split}\label{eq:}
Q_{\bm k_{\rm K} + \delta \bm k} = \frac{\sqrt{3}}{2}e^{i 5\pi/6} (t_1 - 2 t_2)(\delta k_x + i \delta k_y) + O(\delta k^2).
\end{split}\end{equation}
This implies the Dirac speed vanish at $t_1/ t_2 = 2$, which is precisely the ratio realized in the F$^4$ model.

{\it Quantum metric and density fluctuations energy}---While the quantum metric is well-defined for isolated bands, it could become singular at band touching points. Indeed, in our F$^4$ honeycomb model one finds $ g_{ij}(\bm q)  \sim O(1/\delta q^2) $ when $\bm q = \bm q_{\rm K} + \delta \bm q$ is around the K point (and similarly for the K' point). 
More explicitly, we can parameterize $\bm q = \bm q_K + \delta q (\cos \theta, \sin \theta)$ and compute the leading singularity as
\begin{equation}\begin{split}\label{eq:}
g_{\rm K}( \delta \bm q) \approx \frac{1}{\delta q^2} 
\frac{3}{ (2+\sin(2 \theta))^2}
\left(
\begin{array}{cc}
\sin^2( \theta) & -\sin \theta \cos \theta  \\
 -\sin \theta \cos \theta & \cos^2(\theta)
\end{array}
\right).
\end{split}\end{equation}
One might notice that $g_{\rm K}( \delta \bm q) $ has a $\theta$-dependent null vector for all values of $\theta$, but it does not play a role in our analysis.

We can now study how this singularity affects the SMA energy. 
For definiteness, we take $|\Phi \rangle$ to be the state obtained by filling all the negative-energy modes. 
As a technical trick, we further assume a finite linear size $L$ which is not divisible by $3$. This avoids sampling the singular K and K' points in the finite-size momentum mesh and leads to a unique ground state.
The local density operator is given by $\hat n_{\bm R} = \sum_{\sigma= A,B} \hat c_{{\bm R}, \sigma}^\dagger \hat c_{{\bm R}, \sigma}$. In the momentum-space, for $\bm{k} \neq \bm{0}$ and using $\hat f_{\bm q, c}| \Phi \rangle = \hat f_{\bm q, v}^\dagger| \Phi \rangle =0$, we have
\begin{equation}\begin{split}\label{eq:sma_trial}
| {\bm k}\neq \bm{0} \rangle =  \sum_{\bm{ q}} F_{cv}(\bm{k}, \bm{q}) |\bm k, \bm q \rangle,
\end{split}\end{equation}
which can be viewed as a specific superposition of the excited states in Eq.\ \eqref{eq:ph_trial}. 
Here, $F_{cv}(\bm{k}, \bm{q}) $ is a form factor coming from the wave-function overlap $F_{cv}(\bm{k}, \bm{q})  = \bm{u}_{\bm{k+q}, c}^\dagger \bm{u}_{\bm{q}, v}$. The SMA energy for the F$^4$ Hamiltonian is then obtained through the ratio between
\begin{equation}\begin{split}\label{eq:sma_form_factors}
\langle \bm{k}\neq \bm0 |\hat H | \bm{k}\neq \bm0 \rangle =& t \sum_{\bm{q}} \left| F_{cv}(\bm{k}, \bm{q})\right|^2 (e_{\bm{k+q}} + e_{\bm q});\\
\langle \bm{k}\neq \bm0  | \bm{k}\neq \bm0 \rangle =& \sum_{\bm{q}} \left| F_{cv}(\bm{k}, \bm{q})\right|^2. 
\end{split}\end{equation}
As we are ultimately interested in the low-energy excitations, we may specialize to $\bm k \mapsto \delta \bm k$ with $|\delta \bm k| \rightarrow 0$, and then expand the form factor in this limit through 
$\bm{u}_{\bm{q} + \delta \bm k, c} = \bm{u}_{\bm{q}, c} + \sum_i \delta k_i \partial_{q_i} \bm{u}_{\bm{q}, c} + O(\delta k^2)$.
As $\bm{u}_{\bm{q}, c}^\dagger \bm{u}_{\bm{q}, v}=0$, to leading order we have
$\left | F_{cv}(\delta \bm k , \bm q) \right|^2 =  \sum_{i,j} \delta  k_i \delta k_j   g_{ij}(\bm q) + O(\delta k^3)$, where 
\begin{equation}\begin{split}\label{eq:}
 g_{ij}(\bm q) =  \partial_{q_i}(\bm{u}_{\bm{q}, c}^\dagger) \bm{u}_{\bm{q}, v} 
 \bm{u}_{\bm{q}, v}^\dagger   \partial_{q_j} (\bm{u}_{\bm{q}, c} )
 \end{split}\end{equation}
is the quantum geometric tensor for the conduction band in the present two-band problem~\cite{Provost:1980aa,Resta:2011aa,Torma:2022aa, PhysRevLett.131.240001}. 
Here, only the symmetric part of  $g_{ij}(\bm q)$, known as the quantum metric, contributes to $\left | F_{cv}(\delta \bm k , \bm q) \right|^2 $.

The leading term in the $\delta  k  \rightarrow 0$ limit of Eq.\ \eqref{eq:sma_form_factors} is given by $ t \sum_{\bm{q}}  (2 e_{\bm q} )g_{ij}(\bm q)   \delta k_i \delta k_j$. Since $e_{\bm q} \sim O(\delta q^2)$ around the K and K' points, the divergence in the quantum metric is compensated and one finds $\langle \delta \bm k |\hat H | \delta \bm k\rangle \sim   O(1)  \delta k^2$, agreeing with general results derived form a double-commutator bound assuming only the locality and charge conservation symmetry of $\hat H$ \cite{SMA,PhysRevLett.131.220403, arXiv:2405.00785, han2025modelsinteractingbosonsexact}. 
In typical SMA calculations, the normalization factor starts at $O(1)$ and so the SMA energy is gapless and disperses quadratically.

For our model, however, the diverging quantum metric leads to a qualitatively different behavior for the SMA energy. 
The leading behavior of the normalization factor in Eq.\ \eqref{eq:sma_form_factors} is given by 
$\left( \sum_{\bm{q}} g_{ij}(\bm q) \right) \delta k_i \delta k_j$. The sum of $g_{ij}(\bm q)$ over the Brillouin zone is dominated by the contribution from the K and K' points, and its analytic behavior can be extracted by considering the integral
\begin{equation}\begin{split}\label{eq:g_integrated}
\int_{\frac{1}{L}}^{q_0} d \delta q  \, \delta q  \int_{0}^{2\pi} d \theta \,  g_{\rm K}(\delta \bm q)
= \frac{2\pi}{\sqrt{3}} \log \left(q_0 L \right) 
\left(
\begin{array}{cc}
1 &  \frac{1}{2}\\
\frac{1}{2} & 1
\end{array}
\right),
\end{split}\end{equation}
where we introduced a high-momentum cutoff of $q_0 \sim O(1)$ and a low-momentum cutoff of $1/L$.
Note that the matrix in Eq.\ \eqref{eq:g_integrated} is positive definite, and so this leading order term is finite for all directions of $\delta \bm k$. 

\begin{figure}[h]
\begin{center}
{\includegraphics[width=0.4 \textwidth]{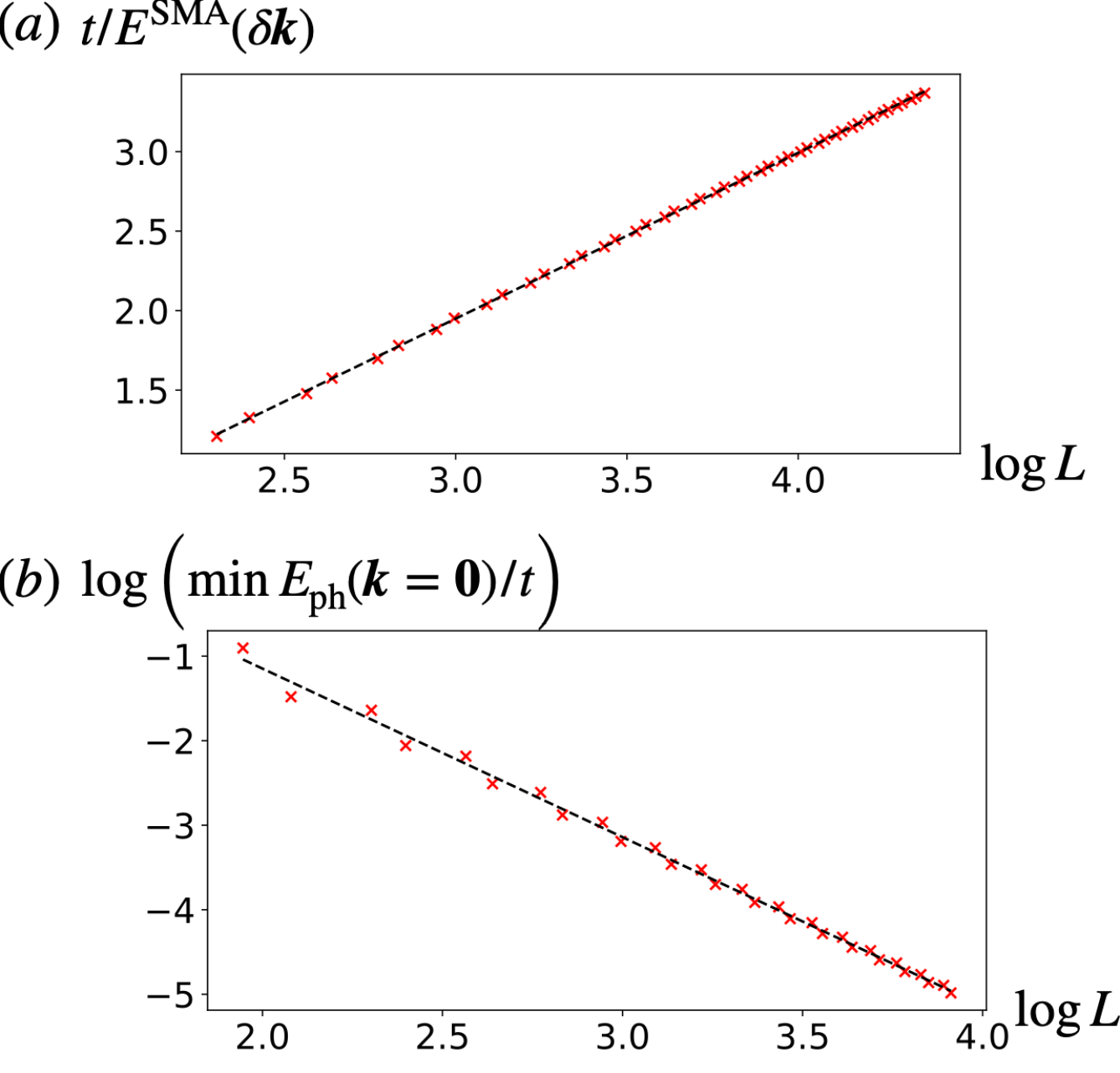}} 
\caption{Fitting of numerical data. Red crosses indicated numerical data and dashed black lines are the linear fits. (a) The linear fit of $t/E^{\rm SMA}(\delta \bm k)$ against $\log L$ gives a slope of $1.04(2)$ and an intercept of $-1.179(7)$. (b) The linear fit of the $\log$ of the minimum particle-hole excitation energy at $\bm k = \bm 0$ against $\log L$ gives a slope of $-1.99(2)$ and an intercept of $2.84(8)$.
\label{fig:fits}
 }
\end{center}
\end{figure}

Therefore, although both the numerator and denominator in the SMA energy start with a quadratic term $\sim \delta k^2$, the coefficient for the numerator is $O(1)$ whereas that for the denominator is $O(\log L)$. This leads to $E^{\rm SMA}(\delta \bm k) \sim O(1/\log L)$ in the limit $\delta k \rightarrow 0$, giving rise to an anomalous gap (Fig.\ \ref{fig:fits}(a)).

{\it Interacting bound of the neutral excitation gap}---
We first Fourier transform the interaction $\hat U$ to find
\begin{equation}\begin{split}\label{eq:}
\hat U = \frac{U}{V} \sum_{\bm q_1,\bm q_2,\bm p} 
\sqrt{
e_{\bm q_2 - \bm p} e_{\bm q_1 + \bm p} 
e_{\bm q_2} e_{\bm q_1} 
}
\hat f^\dagger_{\bm q_2 - \bm p, c} \hat f_{\bm q_2, c}
\hat f^\dagger_{\bm q_1 + \bm p, v} \hat f_{\bm q_1, v},
\end{split}\end{equation}
where $V=L^2$ denotes the volume of the system. The quadratic term acts on the state $|\bm k, \bm q\rangle$ to give 
\begin{equation}\begin{split}\label{eq:}
& \hat f^\dagger_{\bm q_2 - \bm p, c} \hat f_{\bm q_2, c}
\hat f^\dagger_{\bm q_1 + \bm p, v} \hat f_{\bm q_1, v} 
| \bm k, \bm q\rangle\\
= & \delta_{\bm q_2, \bm q + \bm k} f^\dagger_{\bm q_2 - \bm p, c} 
\left( 
\delta_{\bm p, \bm 0} \hat f_{\bm q, v} 
- \delta_{\bm q_1+\bm p, \bm q} \hat f_{\bm q_1, v}
\right) | \Phi \rangle\\
= & \delta_{\bm q_2, \bm q + \bm k} \delta_{\bm p, \bm 0}
 | \bm k, \bm q \rangle
- 
 \delta_{\bm q_2, \bm q + \bm k}\delta_{\bm q_1+\bm p, \bm q} 
 | \bm k, \bm q - \bm p \rangle.
\end{split}\end{equation}
We therefore obtain the matrix elements in the sector with many-body momentum $\bm k$ as
\begin{equation}\begin{split}\label{eq:u_matrix_elements}
\langle \bm k, \bm q' | \hat U | \bm k, \bm q\rangle
=   U e_{\bm q + \bm k  }  \delta_{\bm q', \bm q} 
-   \frac{U}{V} 
\sqrt{
e_{  \bm q'+\bm k  } 
e_{ \bm q + \bm k} 
e_{\bm q'} 
e_{ \bm q } 
}
\end{split}\end{equation}
where we used $\sum_{\bm q_1} e_{\bm q_1} = V$ stemming from the F$^4$ condition.

Next, we consider how the interaction modifies the charge-neutral excitation gap.
While the particle-hole excitations $|\bm k, \bm q\rangle$ are eigenstates of the free-fermion Hamiltonian, they become coupled when the interaction $\hat U$ is added to the Hamiltonian. As translation symmetry is preserved, sectors labeled by $\bm k$ remain decoupled. Let $\hat H^{\rm ph}|_{\bm k=\bm 0}$ be the restriction of the Hamiltonian into the zero-momentum sector, defined by the matrix elements $\langle \bm 0, \bm q| \hat H^{\rm int} | \bm 0, \bm q' \rangle$. 
Fig.\ \ref{fig:ph_energy}(b) plots the low-lying eigenvalues of $\hat H^{\rm ph}|_{\bm k=\bm 0}$ for different system sizes $L$. Fig.\ \ref{fig:fits}(b) shows the $O(1/L^2)$ fit.
We can now find an upper bound to the smallest eigenvalue through
\begin{equation}\begin{split}\label{eq:} 
\min {\rm eig } \left( \hat H^{\rm ph}|_{\bm k=\bm 0} \right) \leq& \min_{\bm q}  \langle \bm 0, \bm q | \hat H^{\rm int} | \bm 0, \bm q\rangle\\
= & \min_{\bm q} \left( 2 t e_{\bm q} + U e_{\bm q} - \frac{U}{V} e_{\bm q}^2 \right)\\
\leq & ( 2 t+ U) \min_{\bm q}   e_{\bm q}  ,
\end{split}\end{equation}
where we used the matrix elements in Eq.\ \eqref{eq:u_matrix_elements} in the second line.

In our finite-size analysis, we take the linear size to be $L=3n+1$ or $L=3n+2$ with integer $n\geq 2$ such that we avoid sampling the K and K' points. We parameterize $\bm q = \bm q_{\rm K}+ \delta \bm q$ with $\delta \bm q = \delta q(\cos\theta, \sin \theta)$. For any fixed $\theta$, one can verify that $e_{\bm q_{\rm K}+\delta \bm q}$ increases monotonically with $0< \delta q \leq 1/7$. In the finite-size sampling of the BZ, there must be at least one sampled momentum within a distance of $1/L$ from the K point. We can therefore bound $\min_{\bm q} e_{\bm q} \leq  \max_{\theta} e_{\bm q_K+ (\cos \theta, \sin\theta)/L} $. In the vicinity of the K point, the single-particle dispersion is maximized at $\theta = 5\pi/4$, and we may evaluate explicitly to find
\begin{equation}\begin{split}\label{eq:}
\min_{\bm q} e_{\bm q} \leq& \eta \left(\delta q = \frac{1}{L}\right);\\
\eta(\delta q) =& 
4 \sin^2 \left( \frac{\pi \delta q}{\sqrt{2}} \right) \left[ \cos \left(   \frac{\pi \delta q}{\sqrt{2}} \right) + \frac{1}{\sqrt{3}} \sin \left(   \frac{\pi  \delta q}{\sqrt{2}} \right)\right]^2
\\
= & 2 \pi^2 \delta q^2 + O(\delta q^3).
\end{split}\end{equation}
Altogether, we obtain the bound
\begin{equation}\begin{split}\label{eq:} 
\min {\rm eig } \left( \hat H^{\rm ph}|_{\rm k=\bm 0} \right) 
\leq & ( 2 t+ U) \eta\left(\frac{1}{L}\right),
\end{split}\end{equation}
and the bound vanishes as $O(1/L^2)$.

\clearpage

\end{document}